       \documentstyle [12pt]{article}
 \evensidemargin\oddsidemargin
\begin{document}
\count0 = 1
 \title{\small{  QUANTUM MEASUREMENT PROBLEM AND SYSTEMS\\ 
SELFDESCRIPTION IN OPERATORS ALGEBRAS FORMALISM \\ }}
\small\author{S.N.Mayburov \\ 
Lebedev Inst. of Physics\\
Leninsky Prospect 53, Moscow, Russia, 117924\\
E-mail :\quad   mayburov@sci.lpi.msk.su}
\date {}
\maketitle
\begin{abstract}
Quantum Measurement problem studied in  Information Theory approach
of systems selfdescription which exploits the information 
acquisition incompleteness for the arbitrary information system.
The studied
 model of  measuring system (MS) consist of measured state S
environment E and  observer $O$ processing input S signal. 
 $O$ considered as the quantum object which interaction with S,E
  obeys to Schrodinger equation (SE).
MS incomplete or restricted states for $O$  derived
by the algebraic QM formalism
 which exploits  Segal and $C^*$-algebras. From 
 Segal theorem for systems subalgebras  it's shown
 that  such restricted states
$V^O=|O_j\rangle \langle O_j|$ describes the classical random 'pointer'
outcomes $O_j$  observed by $O$ in the individual events. 
The 'preferred' basis $|O_j\rangle$ defined by 
$O$ state decoherence via $O$ - E interactions.

\end{abstract}
\vspace{20mm}
\small { \quad 
   Talk given on 'Quantum Information and Computations' workshop, \\}
\small {  \qquad  Triest,  October 2002 }
\vspace{10mm}
\section  {Introduction}

Despite that Quantum Mechanics (QM) is universally acknowledged
physical theory there are still several problems concerned with
its interpretation. Of them the state collapse or objectification
   problem is the most widely
and long discussed ( for the review see $\cite  {Busch}$).
 In our previous  paper \cite {mm}  we  studied
the quantum information transfer  in the measurement process
and resulting from it the information restrictions.
Basing on it we propose the phenomenological formalism
in which  the collapse observation can coexist with
Schrodinger measurement dynamics. In this paper we
reconsider the proposed theory  in Algebraic QM framework
and demonstrate  that it can be derived from C* algebras formalism 
without  any phenomenological assumptions
used in our first paper.

In our approach   
 the information system (Observer) $O$ in the measurement process
 regarded as the
 quantum object which interactions can't be neglected.
Really,  any measurement quantum or classical is 
  the  information acquisition 
 on  the studied system S parameters  by the information system $O$
 via their direct or indirect
interaction. In classical case this interaction can be 
 neglected in the calculations,
 but in QM it can be
  quite important for  obtained measurement outcome \cite {Bre}.
 In our approach
Observer is an information gaining and utilizing system 
(IGUS) of arbitrary structure  $\cite {Gui}$
 which can be  both a human brain or some automatic device 
processing the information. But in all cases it's
the system with some internal degrees of freedom (DF) some  of which
actively interact during
 the information acquisition.
The computer information processing or the perception  by human brain
supposedly corresponds to  the physical objects evolution
 which on microscopic level obeys to  QM laws.
Example  of it are the electron pulses excited in the computer circuits during
the information bits memorization. Such correspondence for the mental
processes isn't proved,  but there are now the strong
experimental evidences that QM can  successfully describe
 a  complex systems including biological ones \cite {Pen}.    
Consequently we  concede in our paper that QM description is applicable
both for  a microscopic    and macroscopic
objects including observer $O$ (he, Bob) \cite {Rov}.
In the simplest case we suppose that
 $O$ state described by  Dirac state vector
$|O\rangle$ ( or density matrix $\rho$ in other cases)
  relative to another  observer $O'$ (she; Alice) \cite {Wig}.

Such approach corresponds with the general trends of quantum information
theory but
our novel point  is that $O$ should  describe
consistently its own quantum state.
Observer selfdescription in the measurement process  (selfmeasurement)
can be  regarded in the context
of the general mathematical  problem of selfreference $\cite {Mitt,Svo}$.
From its study it was argued that $O$ selfdescription
is always incomplete; this results often interpreted as the
analog of Godel Theorem for Information theory \cite {Mitt,Svo}.
 In this framework Breuer demonstrated that $O$ selfmeasurement 
 effects  for classical and quantum measurements 
can be described via the restricted MS states $\cite {Bre}$.
 We'll argue here 
that in quantum case MS restricted states in the measurement
  can be derived from the generalization
of standard QM called algebraic QM which use Jordan or $C^*$- algebras
formalism \cite {Em}.

Here it's necessary to make some  comments on our model premises
and review some terminology.
Our formalism applies both the quantum states in the individual events
- i.e. individual states and the states for quantum ensembles
 called statistical states. Their formal definitions
are given in \cite {Bre} and some of this states features described  below.
 Note only that  for the pure case
the individual and statistical states coincide, yet
 for the mixture they can differ.  
We must stress that throughout our paper the observer consciousness 
 never referred directly and none of their  special  physical properties
beyond standard QM assumed.
 Rather in our model observer $O$ can be regarded as active 
reference frame (RF) which interacts with studied object S
changing $O$ internal state and via it  storing the information about S.
Thus S state description 'from the point of view' of the particular $O$
referred by the terms 'S state in $O$ RF' or simply 'S  state for $O$'.
 The terms 'perceptions', 'impressions' used by us to characterize
   observer subjective description
of experimental  results 
and   defined below in strictly physical  terms \cite {Bre}.
In our theory
the different impressions associated with the different internal
(restricted) $O$ physical
states   which description is operationally unambiguous.

\section {Selfmeasurement  and Quantum States Restrictions}

We start from reminding the main features of our selfmeasurement
theory.
The  microscopic  model   
   of the measuring system (MS)  studied here  includes
the measured state (particle) S and
 observer $O$ which processes and stores the information.
 $O$ interaction with environment E - decoherence
 accounted at the next stage of the model. 
 To illustrate our model  let's consider    
 the measurement   of the observable $\hat{Q}$ of
 S binary  state  performed by $O$: 
$$
     \psi_s=a_1|s_1\rangle+a_2|s_2\rangle
$$
 where $|s_{1,2}\rangle$ are $Q$ eigenstates with eigenvalues $q_{1,2}$.
In our model
$O$ has one effective DF  and its Hilbert space $\cal {H}_O$
contains at least three orthogonal states $|O_i\rangle$
 which are  the eigenstates of $Q_O$ 'internal pointer' observable
with eigenvalues $q^O_i$.
Initial  $O$ state  is  $|O_0\rangle$
and  MS initial state is :
\begin {equation} 
     \Psi^{in}_{MS}=(a_1|s_1\rangle+a_2|s_2\rangle)|O_o\rangle  \label {AAB}
\end {equation}
Let's assume that  S-$O$ measuring interaction starts at $t_0$
and finished effectively at some finite $t_1$. 
MS evolution described by Schrodinger equation (SE) with
 the suitable  S$-O$ interaction Hamiltonian $\hat{H}_I$ 
 which results in
 the final state of MS system :
\begin {equation}
   \Psi_{MS}=a_1|s_1\rangle|O_1\rangle+
a_2|s_2\rangle |O_2\rangle
                                   \label {AA2}
\end {equation}
to which corresponds the density matrix ${\rho}_{MS}$;
 their set denoted $L_q$. 
 $|O_{1,2}\rangle$ are $O$ final states induced by
   the measurement of eigenstates  $|s_{1,2}\rangle$.
 All this states including $|O_i\rangle$ belongs to MS Hilbert space
${\cal H'}_{MS}$ defined in some external $O'$ RF.
 The set (algebra) of all S, $O$ observables denoted
$M_S,M_O$ correspondingly, the set of other (collective) MS observables
which isn't belong to $M_O,M_S$   denoted $M_{MS}$. 
 In practice  IGUS $O$  have many internal DFs, but
 their account  doesn't change principally the results
obtained below.
 We  omit detector D in  MS chain
assuming that S directly interacts with $O$ 
because  the only  D effect is
the amplification of S signal to make it conceivable for O.

 QM predicts  that at time $t>t_1$  for external  $O'$ 
MS is  in the pure state  $\Psi_{MS}$ of (\ref {AA2})
 which is  the superposition of two states for different measurement outcomes.
MS state in $O$ RF $\Psi_{MS}^O$   obtained from
$\Psi_{MS}$ by  the unitary transformation $U'$
and if to consider only MS internal state $\Psi^O_{MS}=\Psi_{MS}$.
 Yet we know  that experimentally   
macroscopic   $O$ observes some  random $Q_O$ value $q^O_{1,2}$
 from which he concludes
that  S final state is $|s_1\rangle$ or $|s_2\rangle$, i.e. 
S state collapses. 
In standard QM  with Reduction Postulate S final  state described by
the statistical ensemble of  individual final states for $O$
described by  the density matrix 
 of mixed state $\rho^s_m$:
\begin {equation}
 \rho^s_m= \sum_i |a_i|^2|s_i \rangle \langle s_i|
                                                              \label {AA33}
\end {equation}
In our model 
we can phenomenologically ascribe to MS the corresponding mixed state :
\begin {equation}
 \rho_m= \sum_i |a_i|^2|s_i \rangle \langle s_i||O_i \rangle \langle O_i|
                                                              \label {AA3}
\end {equation}
 which principally differs from $\rho_{MS}$.
 From $O$ 'point of view' $\Psi^O_{MS}$ describes the superposition of two
contradictory impressions : $Q=q_1$ and $Q=q_2$ percepted simultaneously
which  as Wigner claimed is nonsense \cite {Wig}.
If observer regarded as the quantum object 
 then this  contradiction constitutes famous Wigner 
  'Friend  Paradox' for $O, O'$  $\cite {Wig}$. 
It's quite difficult to doubt both in correctness of
 $O'$ description of MS evolution
by Schrodinger dynamics and in  the state collapse experimental observations.
We attempt to reconcile this two alternative pictures 
in the united formalism which incorporate  both
 quantum   system descriptions 'from outside' by $O'$
and 'from inside' by $O$ - i.e. selfdescription.

In general any classical or quantum measurement 
 of the arbitrary studied system $S'$ is the 
mapping of $S'$ states set $N_s$ on  observer states set $N_O$.  
If an observer $O$ actively interacts with $S'$ so that $O$ state change
can't be neglected then formally $O$ can be regarded as
 the part of the large studied system $S'_T=S+O$ with the states set $N_T$.
  In this approach  $N_O$ is $N_T$ subset and 
$O$ state  is $S'_T$ state projection on $N_O$
 called an restricted  state $R_O$.  
From $N_{T}$ mapping properties the principal restrictions for
MS states discrimination  by $O$ were formulated in
 Breuer theorem \cite {Bre} .
 Namely, if for two arbitrary $S'_T$ 
states $\Phi_{S}, \Phi'_{S}$ 
their restricted  states $R_O, R'_O$ coincide, then for $O$ this $S'_T$ 
states are indistinguishable.
 The origin of this effect  in classical case is easy to 
understand qualitatively : $O$ has less number
 of DFs then $S'_T$ and so can't describe completely $S'_T$ state \cite {Svo}.
 QM introduces additional features connected with observables
noncommutativity and nonlocality which should be also accounted.

For our MS system
it's natural to assume that $R_O$ state depends only of $O$ observables
$M_O$ and doesn't depend of other MS observables.
  Breuer results doesn't permit
to derive the restricted states for arbitrary quantum system
unambiguously and this was our motivation for application of Algebraic
QM for this problem. Breuer proposed  to
 chose as the phenonmenological ansatz
 the partial trace  which for MS state  (\ref {AA2}) is equal to :
\begin {equation} 
   R_O=Tr_s  {\rho}_{MS}=\sum |a_i|^2|O_i\rangle\langle O_i|
      \label {AA4}
\end {equation}
$R_O$ is in fact $\rho_{MS}$ projection into $\cal H_O$ defined
in $O'$ RF and all $R_O$ belong to $\rho_O$  set $L_O$. 


Note that  for MS  mixed state $\rho_{m}$
of (\ref {AA3}) the  corresponding restricted statistical
 state is the same $R^m_O=R_O$.
This equality doesn't mean automatically the
 collapse of MS pure state $\Psi_{MS}$
 because as Breuer argues the collapse presence
   must be  verified for the individual events \cite {Bre}.
 For this purpose it's important to
define the individual mixed state ; as ws noticed for pure MS state it simply
coincides with $\Psi_{MS}$.
 For  the incoming S mixture (\ref {AA3}) of $|S_{1,2}\rangle$
states
  MS individual state objectively exists  in each event $n$, but differs
from event to event  and presented as :
$$
\rho^m (n)=|O_l\rangle \langle O_l|| s_l\rangle\langle s_l|
$$
for random $l(n)$ with probabilistic distribution $P_l=|a_l|^2$ \cite {Bre}.
This individual state can be initially unknown for $O$, but
exists objectively. $\rho^m (n)$  differs from
statistical state (\ref{AA2}) and
its restricted state is $R^m_O(n)=|O_l\rangle \langle O_l|$  also differs 
 from $R_O$ of (\ref {AA4}). Due to it 
 the main condition of Breuer  Theorem violated
 and consequently $O$ can differentiate pure/mixed states 'from inside'
in the individual events. Thus the proposed ansatz doesn't results in 
 the collapse appearance in standard QM \cite {Bre}.
Note hence that the formal states difference in QM by itself doesn't mean
automatically that  $O$ percepts them as different ones.
It should be  at least one observable $A_O \in M_O$ for which $\bar{A}_O$  
expectation values are different which isn't true for Breuer ansatz.

It's  important  to note that even  in this  ansatz 
MS state  for $O$ i.e. 'from inside'  is  different from MS state for
   $O'$ 'from outside' i.e. $\Psi_{MS}$.
 Because of this incompleteness
 $O$ can't see the difference between the
 physically different  pure  states with different $D_{ij}=a_i^*a_j+a_i a_j^*$.
This situation can be called the partial state collapse.
 This states difference reveals
  MS interference term (IT) observable :
\begin {equation}
   B=|O_1\rangle \langle O_2||s_1\rangle \langle s_2|+j.c.
    \label {AA5}
\end {equation}
In standard QM
being measured by $O'$ it gives $\bar{B}=0$ for the mixed MS state (\ref {AA3})
 but  $\bar{B}\neq 0$  for the pure MS states (\ref{AA2}).
Note that $B$ value principally can't be measured by $O$ directly, because
$O$ performs $Q_O$ measurement and $[Q_O,B] \neq0$. 
We define here also $O$ IT observable:
$$
  B_O=|O_1\rangle\langle O_2|+|O_2\rangle \langle O_1|
$$
which will be used below.

Note that formally MS individual state  for $O$ can be written  in doublet 
form $\Phi^B(n)=|\phi_D,\phi_I \gg$, where $\phi_D=\rho_{MS}$ dynamical
 state component, and $\phi_I$ is $O$ information about  MS state
acquired in event $n$. $\phi_I$ is
equal  to $R_O$ for pure state and $\phi_I=R_O^m(n)$ for S mixture
correspondingly.  
 Of course
in this ansatz for pure state $\phi_I$ is just $\phi_D$  projection,
 but in other
selfdescription schemes  they can be more independent
and such  states doublet structure becomes principally important.


The  selfmeasurement theory  permits to discuss the relation between
IGUS evolution and its subjective information (impression)
which supposedly corresponds to $O$ internal state.
Concerning the relations between  the observer state evolution
and his information perception we introduce the following assumptions: 
 for any Q eigenstate $|s_i\rangle$ 
 after S measurement finished at $t>t_1$ 
and $O$ 'internal pointer'state is $|O_i\rangle$ observer $O$
 have the definite impression corresponding  to $q^O_i$ eigenvalue of $Q_O$
that  the measurement event occurred and  the input S state
 is $|s_i\rangle$. This calibration condition for $O$ internal
state is quite nontrivial and important because 
it  settles hypothetically the correspondence between MS quantum
dynamics model and human perception. 
 It also  
related to 'Preferred basis' problem discussed below \cite {Elb}.   
Note that Breuer restricted state $R_O$ corresponds to such 
condition. Futhermore
  if S state is the superposition $\psi_s$  it
supposed that its measurement by $O$ also  results in appearance
for each individual event $n$ of some 
 definite  and unambiguous $O$ impression  denoted as $q^{sup}(n)$. 
This assumption doesn't mean that this impression  for $O$ can corresponds
only  to  one of  mentioned $q^O_i$ eigenvalues but  that it
  devoids of any ambiguities which $O$ quantum state can have.
 In our framework the simplest $O$
toy-model of the information memorization is a hydrogen-like atom
 for which $O_0$ is  its ground state
and $O_i$ are the metastable  levels excited by $s_i$, resulting so into the
final S - $O$ entangled state.

To explain our alternative selfmeasurement theory
consider it first for the considered MS system with
the  initial state (\ref {AAB}).
In its formalism alike in the regarded example
MS state  presented in doublet form $\Phi=|\phi_D, \phi_I\gg$ for dynamical and
$O$ information components. 
The first component of our dual state $\phi_D$ is 
equal to QM density
matrix  $\phi_D=\rho$ and obeys always   
to   Schrodinger-Liouville  equation (SLE)  
 for arbitrary  Hamiltonian $\hat{H}_c$ :
\begin {equation}
    \dot\phi_D=[\phi_D,\hat {H}_c]  \label {AA8}
\end {equation}
which for pure  states is equivalent to Schrodinger equation (SE).
Thus for our MS system $\phi_D=\rho_{MS}$;
 for the information component $\phi_I$ it
 supposed that after S  state $\psi_s$ measurement  in the
individual event $n$  the final state  $\phi_I(n)=V^O$. Here
$V^O=|O_j\rangle \langle O_j|$ is the stochastic state
 with $j(n)$ having  the probabilistic distribution
$P_j=|a_j|^2$.  Thus such doublet individual state $\Phi(n)$ 
changes from event to event and
  $O$ subjective information $\phi_I$ 
is relatively independent of $\rho_{MS}$ and correlated with it only
 statistically.
 Clearly for such restricted states ansatz  $O$ can't differ
the pure and mixed states with the same $|a_i|^2$. Thus Breuer
theorem conditions fulfilled and we'll call
this effect the weak (subjective) collapse.
$\phi_I=V^O_i$ corresponds to $\Psi_{MS}$ branch
 $|\Psi_i\rangle=|O_i\rangle|S_i\rangle$
which describes MS quantum state in $O$ RF i.e. the subjective state.
 An initial state $\phi_D=\rho(t_0)$  defined also by standard QM rules.
 Before measurement starts
 $O$ state vector is $|O_0\rangle$ (no information on S) and
 the  doublet state is
 $\Phi=|\rho(t_0), V_0^O \gg $ where
 $V_0^O=|O_0\rangle \langle O_0|$ is  
the initial $O$ information.

The complete states set in $O$ RF for this individual states is
$N_T=L_q \bigotimes L_ V$ i.e  the  product of dynamical and
information components subsets.
 If we restrict our consideration only on  the
pure states  as done below then $N_T$ is equivalent
  to $\cal {H} \bigotimes \mit L_V$
 and the state vector $|\Psi\rangle$
 can be used as the  dynamical component  $\phi_D$ which 
evolution obeys to SE.
 It assume the
 modification of QM states set, which normally is Hilbert space $\cal H$.
Remind that $\cal {H}$
is in fact an empirical set  choice advocated by fitting QM data.
 Its modifications were 
published already of which most famous is Namiki-Pascazio many Hilbert spaces
 formalism \cite {Nam}.
 Analogous  the  superselection formalism
is well studied in nonperturbative Field theory (QFT)
 with infinite DF number $\cite {Ume}$ and were applied
 for quantum  measurement problem $\cite {Fuk,May2}$.


Of course in this approach
the quantum  states for external $O'$ (and other observers)  also has
the same doublet form $\Phi'$.
 $O'$ doesn't interact with MS and due to it  MS final state for her is
$\phi'_D=\rho_{MS}$ of (\ref {AA2}) and
 $\phi'_I=|O'_0\rangle \langle O'_0|$. Her information
is the same before and after S measurement by $O$, because
$O'$ doesn't get any new information during it. 
 In this approach $O'$ knows that after S measurement
$O$ acquired some definite information $q^O_i$ but can't know without
additional measurements what this information is.
Naturally in this formalism
$O'$ has her own subjective information space $L'_V$ 
   and in her RF  the events states
 manifold is $N'_T=\cal H' \bigotimes \mit L'_V$ for pure states
and the same is true for any number of observers.
 In general if in the Universe
 altogether N observers exists then the complete states manifold
 described in $O$ RF  is 
 $L_T=\cal {H} \bigotimes \mit L_V \bigotimes L'_V ...\bigotimes L^N_V$
 of which only first two subsets elements are observed by $O$ directly. 
 From the described features it's clear that subspace
$L_V$ is principally unobservable for $O'$ (and vice versa for $L'_V,O$),
 because in this formalism only  the measurement
of $\phi_D$ component described by eq. (\ref {AA8}) 
permitted for $O'$.
 Clearly in this ansatz $\Phi, \Phi'$ are unitarily
nonequivalent - i.e. no unitary transformation $U$ can transform
them into each other \cite {Ume}. Note that Selfrefence problem
resolved in this case by use of the natural but unproved
assumption that all observers are similar in relation to
their information acquisition properties.

 For the comparison  in standard QM framework
  S interaction with detector D induces the abrupt and irreversible
S  state  $\psi_s$ change to  random  S state $\psi_j$ - i.e Reduction 
 occurs and in accordance with it D
  pointer acquires definite position $D_j$.
 This process is claimed to be objective, 
i.e. independent of any observer.
Such S,D interaction can't be described by SLE and needs to introduce
 alternative  dynamics ,  which  violates   the
quantum states evolution  linearity and reversibility.
Yet it seems practically impossible to  incorporate in QM
this two contradictive dynamics consistently. In distinction 
in our dual formalism  the dynamical component $\phi_D$
of  MS doublet state evolves linearly and reversibly
in accordance with (\ref {AA8}). This is the objective evolution
in a sense that it described equivalently
relative to any observer.
Only subjective component $\phi_I$ which
 describes $O$ subjective information about $S$ changes stochastically
after S measurement - i.e S,O interaction.


From the doublet individual state $\Phi$ one can derive
 the doublet statistical state for  the quantum ensembles description,
because all the necessary probabilities contained in $\phi_D$.
Due to its importance it's reasonable to define it separately :
$$
|\Theta\gg=|\eta_D,\eta_I\gg=|\rho, R_V\gg 
$$
where $R_V=\sum |a_i(t)|^2 |O_i\rangle \langle O_i|$ is 
the probabilistic mixture of $V^O_i$ states.
$\Theta$  set is $N_{\Theta}=L_q\bigotimes L_R$, where 
$L_R$ set of  diagonal density matrixes with  $Tr\eta_I=1$ but as noticed
above $N_{\Theta}$ is equivalent to $L_q$. 
$|\Theta \gg$ evolution for arbitrary   Hamiltonian
is most simply expressed by the  system of equations
for its components :
\begin {eqnarray}
\frac{\partial{\eta_D}}{\partial{t}}=[\eta_D,\hat{H}_c]    \nonumber \\  
P_j(t)=tr(\hat{P}^O_j \eta_D) \label {CC} \\
\eta_I=\sum P_l |O_l \rangle \langle O_l| \nonumber
\end {eqnarray}
 If $S$ don't interact with $O$ (no measurement)
 then $\eta_I$ is time invariant and MS obeys to the
standard QM evolution for the dynamical component $\eta_D=\rho$
-  statistical state.
Thus our doublet states are important only for measurement-like  processes
with direct  system $ S_A-O$ interactions, but in such case it's  the 
 analog of regarded MS system.
Note that in this theory only $\Phi$ gives complete MS state
description in the individual event, from which its future state
 can be predicted.

The  time of $V^O_0\rightarrow V^O_j$ transition for $O$ is between
$t_0$ and $t_1$  and can't be defined in this  formalism with  the 
larger accuracy, but it doesn't  very important at this stage.
The most plausible assumption compatible
with standard QM is that $O$ perception time $t_p$ in
MS measurement has the distribution :
$$
  P_p(t)=c_p \sum \frac{\partial P_i(t)}{\partial t}\quad ; \quad i\ne0
$$ 
where $c_p$ is normalization constant. Note that this result
 is compatible with standard QM.

The preferred basis (PB) problem importance
is acknowledged in quantum measurement theory
\cite {Busch}. In its essence, $\Psi_{MS}$ decomposition on $O$,S states
in general isn't unique and so any theory must explain why namely
 $|O_i\rangle$ states appears in final mixture $\rho^m$. 
In our theory PB acquires the additional aspects being related
to $O$ information recognition. As indicated above 
we choose as the
 calibration condition that $|O_i\rangle$ state percepted by $O$ as
 $q^O_i$ value corresponding to $q_i$ value of S parameter $Q$.
 But it's not clear why  such states set responds to it
and not some other  $|O^C_j\rangle$  - eigenstates of some $Q^C_O$,
belonging to another orthogonal basis. For example, it can be
$|O_{\pm}\rangle= \frac{|O_1\rangle \pm |O_2\rangle}{\sqrt{2}}$ 
for the binary subspace ${\cal H}_O$.
Yet the situation changes principally, if to account decoherence - i.e.
$O$ interaction with environment E
 $\cite {Zur,Gui}$.  In the simple 
decoherence model E considered here
consists of $N$ two-level systems (atoms) independently 
interacting with $O$ with $H_{OE}$ Hamiltonian, which for arbitrary
E states $|E^0\rangle$  at large $t$ gives:
 $|O^E_i\rangle |E^0\rangle  \rightarrow|O^E_i\rangle |E^0_i\rangle$
, where $|O^E_i\rangle$ belongs to an arbitrary orthogonal basis $O^E$
of $O$  states.  $|E^0_i\rangle$ are E  states which aren't necessarily
 orthogonal; S,$O$,E joint Hilbert space denoted ${\cal H}_{MS+E}$.
 Tuning specially the  measurement
Hamiltonian $H_{I}$ one can make two basises equal:
  $|O^E_i\rangle=|O_i\rangle$ and  only this case will be regarded here.
If in S measurement at $t<t_1$
 $O-$E interaction can be neglected then
under simple assumptions it results in the final MS-E state :
\begin {equation}
    \Psi_{MS+E}=\sum a_i|s_i\rangle|O_i\rangle|E^0_i\rangle
                                             \label {DD1}
\end {equation}
It was proved that such triple decomposition is unique, even
if $|E_i^O\rangle$ aren't orthogonal \cite {Elb}.
 Thus PB problem formally resolved if the decoherence accounted and
this is essential also  for our model. It isn't necessary
in our case to use $ N\rightarrow \infty$ limit, E can be also  a finite,
closed system.

 In addition the  decoherence results in the
important consequences for the definition of the  mentioned information basis.
Really the memorized states $|O^C_i\rangle$ excited by $S_i$ signals
 must be stable or at least long-living. But as follows from eq. (\ref{DD1})
 any  state $|O^C_j\rangle$ different from one of $|O_i\rangle$
in the short time would split into $|O_i\rangle$ combinations - entangled
$O$,E  states superposition. 
But  our calibration condition  demands that at least
$Q_O$ eigenstates will be conserved copiously and not transferred
to any combinations.
Thus in this model  $O$-E decoherence interaction selects
 the basis of long-living
$O$ eigenstates which supposedly
describes $O$ events perception and
memorization, i.e. $\phi_I$.
It means that if our S signal is $Q$ eigenstate
 inducing $Q_O$ eigenstate $|O_i\rangle$ then it's memorized
by $O$ for long time. As the result the only $O$ observable
which can be memorized by $O$ is $Q_O$ ; for any other $O$ observable
$Q'_O$ and any $\Psi_{MS}$ one obtains that in MS,E final state
$\bar{Q}'_O=0$.

Decoherence influence should be accounted in any 
 $O$ selfdescription formalism.
In Breuer ansatz $O$ interaction with E states accounted
analogously to  S states, so that
$R_O=Tr_{S,E}\rho$ derived taking trace both on $S$ and $E$ states.
 For our dual formalism
 an analogous approach permits to derive $| \Theta \gg$, thus defines 
 $\phi_I$ distribution. In the individual events $\phi_I$
correspons to $|O_i\rangle|S_i\rangle|E^O_i\rangle$ branch.
 In other aspects
 decoherence doesn't change our selfmeasurement model.


\section { System  Selfdescription in Algebraic QM}

Now we  regard  the quantum system selfmeasurement 
in  Algebraic QM framework  and derive MS restricted states
 $C^*$ algebras  formalism \cite {Bra}.
 We'll show that algebraic QM 
applied to  the systems selfdescription  in fact corresponds to our
phenomenological   results.  
Algebraic QM  includes the  unitary nonequivalent representations
of commutation relations which  describes successfully
phase transitions and other nonpreturbative effects 
which  standard QM fails to incorporate. 
Consequently we have the serious premises to regard Algebraic QM
as the consistent generalization of standard QM.
As noticed above the nonperturbative QFT formalism
 was applied also to  the study of measurement 
problem \cite {May2}. Application of Operators Algebras 
to quantum measurements studied also in \cite {Pri}
 but without the  selfdescription consideration.

Remind that in standard QM the primordial structure is the states set
- Hilbert space $\cal H$ on which an observables - Hermitiam operators
defined.
In distinction in Algebraic QM formalism as the fundamental structure 
 is the  observables algebra $\cal{U}$
which can be Jordan or Segal algebra with
formal  extension to $C^*$-algebra \cite {Em}. Consequently
the system $S_f$   states set $\Omega$ properties defined by  $\cal{U}$
and under some assumptions $\Omega$ is the  vector space dual to $\cal U$
defined  by the famous GNS construction. The obtained algebraic states
 $\varphi \in \Omega$
are equivalent to the normalized positive 
linear functionals on $\cal U$.
 Formally in the situations described by standard QM they
corresponds   to QM density matrixes $\rho$.
The pure quantum states $\varphi$ in this formalism
 are an  extremal $\Omega$ points.

  In many realistic situations for
given measurement set-up not all  the system $S_f$ observables are available 
for experimentalist, but only some restricted subset to which
corresponds $\cal U$ subalgebra $\cal{U}_R$. It was shown that for some 
$\cal {U}_R$
 the corresponding states set $\Omega_R$ can be constructed
which states $\varphi_R$  differs from $\varphi$  \cite {Em}.
For this case Segal theorem demonstrates that if all $A_R\in \cal{U}_R$
commute and unit operator $I\in \cal {U}_R$,
 then  $\Omega_R$ is set of classical states $\varphi_R$
with  no superpositions between them \cite {Seg}.
In Algebraic QM notations $A_R$ expectation values defined as
 $$
    \bar{A}_R=\langle \varphi;A_R\rangle=\langle \varphi_R;A_R\rangle
$$
 coincide for $\Omega, \Omega_R$ states.
$A_R$ measurement corresponds to the simultaneous measurement
of all $A_R$ projectors $P^A_i$ which expectation values 
$\bar{P}^A_i$ reproduces the probabilistic $A_R$ distribution.


For our MS,E system
 $\cal{U}$ Segal algebra for  observables sets $M_S,M_O,M_{MS},M_E,...$
 can be defined and
 $\varphi$ states set $\Omega$ 'spanned' on them
  naturally  is equivalent to ${\cal H}_{GS}$.
In this case $O$ complete observables set is $M_O$ but 
 as was demonstrated above for
the regarded MS,E measurement dynamics $O$
can measure (percept) only the sinlge observable $Q_O$ 
 which illustrated by $\Psi_{MS+E}$ of (\ref {DD1}). So in this case
$\cal{U}_R$ subalgebra consists of $Q_O$ and $I$ which obviously is available
 for $O$.   Then from Segal theorem 
 MS restricted   states $\varphi_R$
  should be the discrete classical states.
 If  the incoming S state
$\psi_s$ is S eigenstate $|S_i\rangle$ then from our calibration condition
 such MS  restricted states can be associated with our 
$O$  states $\varphi^O_i \sim |O_i\rangle\langle O_i|$
to which corresponds $\phi_I=V^O_i$ of our dual formalism. 
Thus in any such individual S measurement event 
 the final MS restricted
 state  from  $O$ 'point of view' describes definite discrete
$O_i$ value so that $\bar{Q}_O=q^O_i$.
 But if  the incoming $\Psi_{MS}^{in}$ described by (\ref{AAB}) with
several  $a_j\neq 0$   
i.e. is $Q$ eigenstates superposition one obtains :
$$
 \bar{Q}=\bar{Q}_O=\langle\varphi_R;Q_O\rangle=
\langle\varphi;Q_O\rangle=\sum |a_i|^2 q^O_i
$$
and no $\varphi^O_i$ state corresponds to such $\bar{Q}_O$.
To interpret this result note first that for the incoming S mixture 
 of $|s_i\rangle$ eigenstates (\ref {AA2}) the final statistical
 MS state described for
$O$ as the mixture $\rho^m_O=\sum |a_i|^2 \varphi^O_i$
 with the same expectation value $\bar{Q}_O$.
$q^O_i$ probabilistic distribution naturally described by $P_i=|a_i|^2$.
As was shown above the difference between the pure and
mixed MS states reflected by $B$ IT expectation values.
Thus
 $O$ observation of S pure/mixed states difference means
that $O$ acquires some information on MS  $B$  expectation value.
 But $B \notin \cal {U}_R$ and isn't correlated with $Q_O$ via
some interaction alike $Q$ of S; 
so this assumption  for $\varphi_R$ states
results in contradiction.
Summing all the given arguments we conclude
that for the pure incoming S state $\psi_s$
 $O$ observes stochastic $q^O_i$ distribution with 
probabilities $|a_i|^2$ described by statistical mixture
 $\varphi_R=\sum |a_i|^2\varphi_i^O$ and thus can't be discriminated
from incoming S mixture with the same $P_i$.
It corresponds to the natural generalization of Algebraic QM premises
settling that not only statistical but individual restricted
states defined solely by ${\cal U}_R$.
Alike in the previous calculations
 we assume  that the individual S mixed states  
described by one of pure states $|S_i\rangle$ which after S measurement
induces the corresponding restricted state $\varphi^O_i$.
Our doublet state $\Phi$ components $\phi_D,\phi_I$
 corresponds to $\varphi$ and
$\varphi^O_i$ in the individual event.


 In fact Algebraic QM supports
the simple and consistent selfmeasurement picture : any $O$
can observe only the object parameters for which it posses the suitable
 'internal' instrument. For example $O$ can't observe $|O_i \rangle$  
superpositions because $O$ detecting structure doesn't permit
IT observable $B_O$ to be measured simultaneously with $Q_O$.
 Such considerations
were already regarded above in the phenomenolgical picture of our
 dual formalism but in Algebraic QM they acquire additional support.
Restricted subalgebras $\cal{U}_R$  were applied already
for the information restrictions  stipulated by 
 the practical impossibility of 
 the large  systems complete description \cite {Em}. 
We apllied them for a restrictions induced not by a studied system, but
 by IGUS $O$ quantum structure,
yet we  don't see any contradictions in such generalization.
Note that in Algebraic QM an appearence of stochastic final states 
related to the phenomena of spontaneous symmetry breaking. By the
analogy the effect of measurement results randomization can be
called Information symmetry breaking. Its origin connected
with the fact that  the final physical
states $\varphi_i^O$ doesn't posses the symmetry of the incoming measured state
$\Psi^{in}_{MS}$ and the final $q^O_i$ values are undecidable
 according to Svozil approach \cite {Svo}.
We don't discuss here Algebraic QM foundations consistency
and in particular the feasilbility of the algebraic states
which deserve the separate consideration \cite {Pri2}.

We perform algebraic calculations for our very simple
IGUS $O$ model, which probably is more primitive then any realistic
 IGUS structure.
But  in Algebraic QM 
the only important condition for the classicality appearence
is $\cal{U}_R$ observables commutativity and it's reasonable to expect
 it to be feasible
also for complex IGUS structures. Clearly this subalgebras
properties can't depend directly on the surrounding E properties
and its particular state. Thus we can suppose that obtained results
can be true not only for closed but also for the open  systems.

\section {Discussion}

Our studies demonstrates that the account of the information
system quantum properties permit to explain the state collapse
as the consequence of  the principal  system
selfdescription incompleteness.  It was studied  with the simple IGUS
$O$ model in which $O$ states decoherence selects the preferred
'perception' basis.
Breuer selmeasurement study shows that
by itself  the inclusion of observer as the quantum object
into the measurement scheme doesn't result in the
 collapse appearance \cite {Bre}.
Our theory indicates that to describe the collapse it it's necessary also
 to modify the quantum states set.
 In this
  approach different  observers 
 becomes nonequivalent
in a sense that the  physical reality description becomes principally different
for each of them \cite {Rov}. This nonequivalence
reflected by the presence of subjective component $\phi_I$ available directly
only for particular $O$ and for him the subjective state collapse can be
obtained. This theory  permit to conserves
Schrodinger evolution for arbitrary quantum system.
 Eventually the consistent selfmeaurement formalism 
constructed alternative to Breuer  formalism.

 It was shown here  that Algebraic QM formalism 
presents the additional arguments in favour of our selfmeasurement
theory. It permits to calculate
  the restricted $O$ states $\varphi_R$ from 
 the given $O$ physical structure and corresponding $O$ observables
algebra.
 Obtained $O$ states corresponds to our phenomenological doublet
states $\Phi$ and are unitarily nonequivalent for different
observers.
 Algebraic QM formalism
- Jordan , Segal and $C^*$-algebras of operators is acknowledged
generalization of standard QM. In this paper it was applied for 
the simplest measurement model but if this formalism universality
will be proved it will mean that the proposed selfmeasurement
 theory follows from the established Quantum Physics realm
without any additional assumptions \cite {Em}.

From this considerations
the natural question arise : does the observation of random
outcomes $\phi_I=V^O_j$ means that before the measurement starts 
 S state can be characterized by some objective  'hidden parameter'
$j_S$ ? Our theory is principally different from Hidden Parameters
theories where this stochastic parameters influence the quantum state
dynamics and  so differs from standard QM. Due to it in our 
 theory  $\phi_I$ internal parameter $j$ can be 'generated' 
during S-$O$ interaction and don't exists objectively before it starts.
This is analog of spontaneous symmetry breaking effect
derived in Algebraic QM.
Yet our dual theory demonstrates that the probabilistic realization
 is generic and unavoidable for QM and without it QM supposedly can't
acquire any operational meaning. Wave-particle dualism
was always regarded as characteristic QM  feature, but in our theory
it has straightforward  correspondence in dual 
seldmeasurement formalism.

The ideas close to our dual theory were  discussed in QM modal
interpretation, but they have there quite unclear
philosophical motivations \cite {Busch}. Now this is the whole class
of different theories, of which the most close to us 
is  Witnessing interpretation by Kochen $\cite {Koch}$.
His theory phenomenologically supposed that for apparatus $A$
measured value $S$ in pure state always has random definite value
$S_j$, yet no physical arguments for it and no mathematical formalism differ
from standard QM  were proposed.

In general all our experimental conclusions are based on human
subjective perception. Regarded computer-brain perception analogy
in fact means that the human signal perception also defined by $\bar{Q}_O$
values. Despite that this analogy looks quite reasonable we can't
give any proof of it. 
We present here very simple measurement theory and we don't regard
it as the final solution of measurement problem. Yet from its results
we believe that it's impossible to solve it without account of 
$O$ interaction with  the measured system at quantum level \cite {Zur2}.
 We must  stress that our theory doesn't need any
addressing to  human observer consciousness. Rather in this model 
$O$ is active RF which internal state excited by the interaction
with the studied object.  
This approach to the measurement problem 
has much in common with Quantum reference frames theory introduced by
 Aharonov  $\cite {Aha3}$.

Historically the possible influence of observer on measurement
process was discussed first by London and Bauer \cite {Lon}. They supposed
that  Observer Consiousness (OC) due to 'introspection
action' violates in fact Schrodinger equation for MS 
and results in state reduction. This idea was seriously  criticized  
by Wigner \cite {Wig}. In distinction in our dual theory OC perception     
doesn't violate MS Schrodinger evolution from $O'$ point of view.
But measurement  subjective perception in it also performed by OC
and its results partly independent of dynamics due to
its dependence on the stochastic component $\phi_I$.
 This effect deserves further
discussion, but we believe that  such probabilistic behavior
is general IGUS property not related to OC only.


\end {document}